\documentclass[useAMS,usegraphicx,usenatbib]{mn2e}
\usepackage{subfig}
\usepackage{aas_macros}

\newcommand{\iso}[2]{\ensuremath{^{#1}\rm{#2}}}
\newcommand{\msun}{\ensuremath{M_{\odot}}}
\title[Overshoot mixing in Nova - the underlying composition]
  {Convective overshoot mixing in Nova outbursts - The dependence on the composition of the underlying white dwarf}
\author[Ami Glasner et al.]
  {S. Ami ~Glasner,$^1$ 
   Eli ~Livne,$^1$ James W.~Truran$^2$
\newauthor
\\
  $^1$Racah Institute of Physics, The Hebrew University, Jerusalem 91904,  Israel\\
  $^2$Department of Astronomy and Astrophysics, Enrico Fermi Institute, \\
   and Joint Institute for Nuclear Astrophysics, University of
   Chicago, \\
   5640 South Ellis Avenue, Chicago, IL 60637}

\date{Released 2012 June 28}

\pagerange{\pageref{firstpage}--\pageref{lastpage}} \pubyear{2012}

\def\LaTeX{L\kern-.36em\raise.3ex\hbox{a}\kern-.15em
    T\kern-.1667em\lower.7ex\hbox{E}\kern-.125emX}

\begin{document}

\label{firstpage}

\maketitle

\begin{abstract}

 We present here, for the first time, a 2D study of the overshoot convective mechanism 
 in nova outbursts for a wide range of possible compositions of the layer underlying the accreted envelope.
 Previous surveys studied this mechanism only for solar composition matter accreted on top of carbon
 oxygen (C-O) white dwarfs. Since, during the runaway, mixing with carbon enhances the hydrogen burning rates dramatically, one should
 question whether significant enrichment of the ejecta is possible also for other underlying compositions (He, O, Ne, Mg), predicted by stellar
 evolution models. We simulated several non-carbon cases and found significant amounts of those underlying materials in the
 ejected hydrogen layer. Despite large differences in rates, time scales and energetics between the cases, our results show that
 the convective dredge up mechanism predicts significant enrichment in all our non-carbon cases, 
 including helium enrichment in recurrent novae. The results are consistent with observations.

\end{abstract}


\section{Introduction}

\label{intro}

 Almost all classical and recurrent novae for which reliable abundance determinations exist show enrichment 
(relative to solar composition) in heavy elements and/or helium. It is now widely accepted that the source
for such enrichment is dredge up of matter from the underlying white dwarf to the accreted envelope.
A few mechanisms for such mixing were proposed to explain the observations :
 Mixing by a diffusion layer, for which diffusion during the accretion phase builds a layer 
 of mixed abundances (\cite{pk84,kp85,IfMc91,IfMc92,FuI92}); Mixing by shear instability induced by differential rotation during the accretion phase 
 \citep[]{Du77,kt78,Macd83,lt87,ks87,ks89}; Mixing by shear gravity waves breaking on the white dwarf surface in which a resonant interaction between 
  large-scale shear flows in the accreted envelope and gravity waves in the white dwarf's core can induce mixing of heavy elements into the envelope
 \citep{Rosner01,alex02,Alex2D} and finally - Mixing by overshoot of the convective flow during the runaway itself 
 \cite{woo86,saf92,sha94,gl95,glt97,Ker2D,Ker3D,glt2007,Cas2010,Cas2011a,Cas2011b}. 

 In this work we focus on the last of these mechanisms that proved efficient for C-O white dwarfs. 
Mixing of carbon from the underlying layer significantly enhances the hydrogen burning rate. The enhanced 
burning rate drives higher convective fluxes, inducing more mixing \cite{gl95,glt97,glt2007,Cas2010,Cas2011a,Cas2011b}. 
Therefore, the fact that the underlying layer
is rich in \iso{12}C is a critical feature of all the overshoot convective models that have been analyzed up to this work. 
According to the theory of stellar evolution for single stars, we expect the composition of the underlying
white dwarf to be C-O for masses in the range  $0.5-1.1 M\sun$  and ONe(Mg) for more massive white dwarfs \cite{GpGb01,dom02,Gb02}.
Observations show enrichment in helium, CNO, Ne, Mg and heavier elements \citep{sst86,Gehrz98,Gehrz08,Ili02}. For recurrent novae, helium 
enrichment can achieve levels of $40-50\%$ \citep{web1987,anu2000,diaz2010}. 
High helium abundances can simply be explained as 
the ashes of hydrogen burning during the runaway \citep{Hernanz2008}, but one can not exclude the possibility that the source of 
He enrichment is dredge up from an underlying helium layer.    
We therefore found it essential to study nova outbursts for which the composition of the underlying layer is different from C-O. 
The models studied here extend the work we publish in the past. 
As a first step we study here the runaway of the accreted hydrogen layer on top of a single white dwarf, changing only its composition. 
Having a fixed mass and radius we can compare the timescales, convective flow, energetics and dredge up in the different cases.  
A more comprehensive study which varies the white dwarf's mass with compositions is left to future work (CO, ONe(Mg) or He rich).

The present study is limited to 2D axially symmetric configurations. The well known differences between 2D and 3D unstable flows can
yield uncertainties of few percents on our results, but can not change the general trends, as previous studies showed reasonable agreement between 
2D and 3D simulations with regard to integral quantities, although larger differences persist in the local structure ( \cite{Cas2011b}).
We therefore regard our present results as a good starting point to more elaborated 3D simulations. 
In the next section we describe the tools and the setup of the simulations. In subsequent sections we describe the results for each
initial composition and then summarize our conclusions.

\section{Tools and initial configurations}
\label{tools}

All the 2D simulations presented in this work start from 
a  1D hydrostatic configurations, consisting of a $1.147 M\sun$ CO core with an outer layer of 
 $~10^{-4}\,\msun$ composed of CO, ONe(mg) or helium, according to the studied case as we explained in the introduction.
The original core was built as a hydrostatic CO politrop that cooled by evolution to the desired central temperature ($2\times10^7 {\rm K}$). 
The 1D model evolves using Lagrangian coordinates and does not include any prescription for mixing at the bottom of the envelope.
The core accretes matter with solar abundance and the accreted matter is compressed and heated.
Once the maximal temperature at the base of the accreted envelope reaches a temperature of  $9\times10^7 {\rm K}$, 
the whole accreted envelope $3.4\times10^{-5}\,\msun$ and most of the underlying zone $4.7\times10^{-5}\,\msun$ , 
are mapped onto a 2D grid, and the simulations continue to runaway and beyond using the 2D hydro code VULCAN-2D \cite{liv93}.
This total mass of   $8.1\times10^{-5}\,\msun$ is refered as the total computed envelope mass. 
Using same radial zoning in the 1D grid and in its 2D counterpart, the models preserve hydrostatic equilibrium, accurate to better than
one part in ten thousand. Since the configurations are unstable, non radial velocity components develop very quickly from the small
round-off errors of the code, without introducing any artificial initial perturbation.

Further computational details of the 2D simulations are as follows.
The inner boundary  is fixed, with assumed zero inward luminosity. The outer boundary follows the expansion of the envelope taking advantage 
of the arbitrary Lagrangian-Eulerian (ALE) semi-Lagrangian option of the VULCAN code whereas the burning regions
of the grid at the base of the hydrogen rich envelope are purely Eulerian. More details are presented in Glasner et al. (2005, 2007).
The flexibility of the ALE grid enables us to model the burning zones at the bottom of the hydrogen rich envelope with
very delicate zones in spite of the post-runaway radial expansion of the outer layers.
The typical computational cell at the base of the  envelope, where most of the burning takes place, is a rectangular cell with 
dimensions of about $ 1.4 km \times 1.4 km $ . Reflecting boundary conditions are imposed at the lateral boundaries of the 
grid. Gravity is taken into account as a point source with the mass of the core, and the self gravity of the envelope is ignored.
The reaction network includes 15 elements essential for the hydrogen burning in the CNO cycle, it includes the isotopes:
 \iso{1}H, \iso{3}He, \iso{4}He  \iso{7}Be, \iso{8}B, \iso{12}C,  \iso{13}C, \iso{13}N, \iso{14}N, 
\iso{15}N, \iso{14}O,  \iso{15}O,  \iso{16}O , \iso{17}O and \iso{17}F.

\section{Results}

 In order to compare with 1D models and with previous studies we present here five basic configurations: 

  1) The outburst of the 1D original model without any overshoot mixing.

  2) An up to date model with an underlying C-O layer.

  3) A model with a Helium underlying layer.  

  4) A model with an underlying O(Ne) layer.

  5) A toy model with underlying \iso{24}Mg. 
  
This model demonstrates the effects of possible mixing of hydrogen with $^{24}$Mg on the runaway. (In a realistic model the amounts of  \iso{24}Mg in the core are expected to be a few percents (\cite{GpGb01,siess06}), but higher amounts can be found in the very outer layers of the core (\cite{berro94} ).

 The computed models are listed in  Table~\ref{tab:models}. 
 In the next sections we present the energetics and mixing results for each of  the models in the present survey.
 
\begin{table}
\caption{Parameters of the Simulated Initial Configurations}
\label{tab:models}
 \begin{tabular}{@{}llccc}
  \hline
  Model & $Underlying$ & $T_{max}$
        & $Q_{max}$
        & $remarks$ \\
  \hline
 m12   &  -    &    $2.05 $ &    $ 4.0 $      &  1D   \\
 m12ad &  CO   &    $2.45 $ &    $ 1000.0 $   &   -     \\
 m12ag &  He   &    $  -  $ &    $  -  $      &   -    \\
 m12dg &  He   &    $  -  $ &    $  -  $      &    $T_{base}=1.5\times10^8 K$  \\
 m12al &  O    &    $  -  $ &    $  -  $      &   -    \\
 m12bl &  O    &    $  -  $ &    $  -  $      &    $T_{base}=1.05\times10^8 K$   \\
 m12cl &  O    &    $  -  $ &    $  -  $      &    $T_{base}=1.125\times10^8 K$   \\
 m12dl &  O    &    $2.15 $ &    $ 82.4  $    &    $T_{base}=1.22\times10^8 K$   \\
 m12kk &  O    &    $  - $    &   $  -  $     &  \iso{24}Mg rates   \\
 m12jj &  O    &   $2.46 $  &    $1000.0 $    &  \iso{24}Mg rates+ $T_{base}=1.125\times10^8 K$   \\
\hline
\end{tabular}
\medskip
$T_{max}$; maximal achieved temperature [$10^8$ Kelvin].                
$Q_{max}$; maximal achieved energy generation rate [$ 10^{42} erg/sec$]
\end{table}

\subsection{The 1D original model}

 The initial (original) model, composed of a  $1.147 M\sun$ degenerate core, accretes  hydrogen rich envelope 
 at the rate of $1.0 \times 10^{-10} \msun/year$.
 When the temperature at the base of the accreted envelope reaches  $9\times10^7 {\rm K}$ the accreted mass is $3.4\times10^{-5}\,\msun$.
 We define  this time as t=0, exceptional test cases are commented in Table~\ref{tab:models}.
 Convection sets in for the original model a few days earlier, at $t=-10^6 sec$, when the base temperature 
 is  $3\times10^7 {\rm K}$.
 The 1D convective model assumes no overshoot mixing; therefore, convection has an effect only on the heat transport and abundances within the convective zone.
 Without the overshoot mixing burning rates are not enhanced by overshoot convective mixing of CO rich matter. As a reference for the 2D simulations, 
 we evolved the 1D model all through the runaway phase.
 The time to reach the maximal energy production rate is about $ 2400 sec $, the maximal achieved temperature is $2.05\times10^8 K$,
 the maximal achieved burning rate is   $4.0\times10^{42} erg/sec$ and the total nuclear energy generated up to the maximum 
 production rate (integrated over time) is $0.77\times10^{46} erg $.

\subsection{The C-O underlying layer}
\begin{figure}
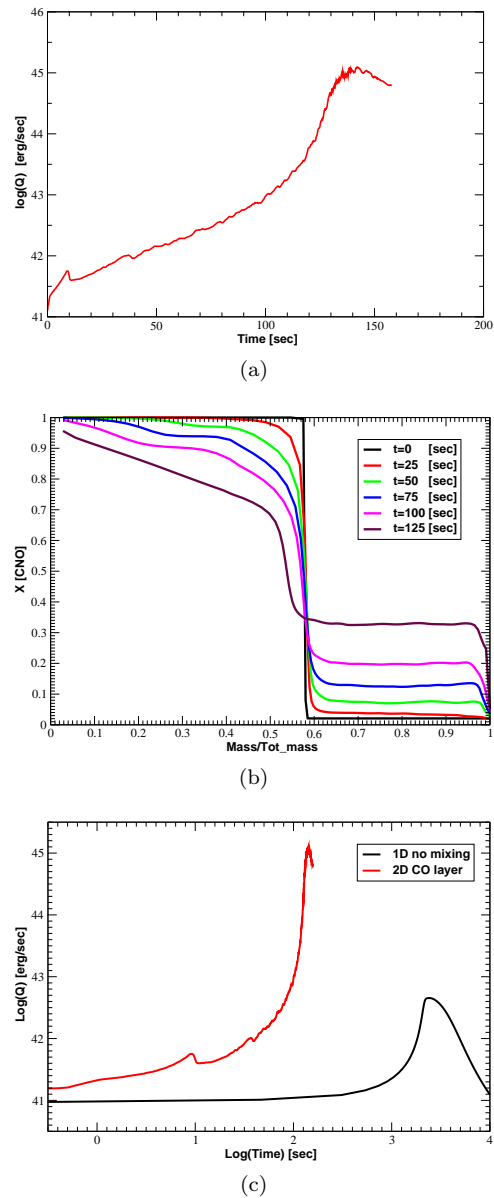

\centering

        \hspace{1.0in}
        \subfloat[]{\includegraphics[clip,width=.36\textwidth]{v-q-m12ad.eps}}

        \subfloat[]{\includegraphics[clip,width=.36\textwidth]{v-x-m12ad.eps}}

        \subfloat[]{\includegraphics[clip,width=.36\textwidth]{v-q-1D-CO.eps}}

 \caption{The CO model
 a - the logarithm of the energy generation rate as function of time, b- the 
 total abundance of the CNO elements as function of the fractional mass coordinate of the computed envelope (see section 2) at various times
 that are indicated on the plot,
 c - the log of the total energy generation rate as function of the logarithm of the time coordinate for the 1D model and for the 2D CO model.
  } 
 \label{co}

\end{figure}

  We summarize here the main results of the underlying C-O case (computed already in \cite{glt2007} and repeated here),
  which is the most energetic case. A comparison of the history of the burning rate (Fig.~\ref{co}) with Figure 3 in \cite{glt2007} confirms that our current numerical results agree with those of our earlier publication. 
  In  this figure (Fig.~\ref{co}) we also present the amount of mixing at various stages of the runaway.
  The main effects of the convective underlying dredge up are:

\begin{itemize}

\item The convective cells are small at early stages with moderate velocities of a few times  $ 10^{6} $ cm/sec.
 As the energy generation rate increases during the runaway, the convective cells merge and become almost
 circular. The size of the cells is comparable to the height of the entire envelope, i.e. a few pressure scale heights.
 The velocity magnitude within these cells, when the burning approaches the peak of the runaway, is a few times $ 10^{7} $ cm/sec.

\item  The shear convective flow is followed by efficient mixing of C-O matter 
 from the core to the accreted solar abundant envelope.  The amount of C-O 
 enrichment increases as the burning becomes more violent and the total amount of mixing is above $30\%$ (Fig.~\ref{co}). 
 
\item Mixing induces an enhanced burning rate, relative to the non mixing 1D case, by more than an order of magnitude. 
 The maximum rate grows from $  4.0\times10^{42} erg/sec $ to  $ 1000.0\times 10^{42} erg/sec $. 
 The enhanced rates raise the burning temperature and shorten the time required to reach  
 the maximal burning rate. The maximal achieved temperature increases from  $ 2.05\times 10^8 K$ 
 to  $ 2.45\times 10^8 K$ and the rise time to maximum burning decreases from $ 2440 sec $ to $ 140 sec $. 
 The total energy production rates of the 1D and the 2D simulations are given in Fig.~\ref{co}.
 The enhanced burning rate in the 2D case will give rise, at later stages of the outburst, to an increase in the kinetic energy of the ejecta. 
 Unfortunately, since the hydro solver time step is restricted by the Courant condition we can not run the 2D models through to the coasting phase.
 A consideration of  the integrated released energy  at the moment of maximal burning rate reveals that the burning energy grows from $ 0.77\times 10^{46} erg$ in the 1D model to $ 1.45\times 10^{46} erg$ in the 2D model, a factor of 2.

 \end{itemize}

 Another interesting feature of the 2D C-O simulations is the appearance of 
 fluctuations, observed during the initial stages of the runaway (Fig.~\ref{co}). Such fluctuations are not observed in the 1D model. 
 The fluctuations are a consequence of the mixing of the hot burning envelope matter with the cold underlying white dwarf matter.
 The mixing has two effects. The first is cooling as we mix hot matter with cold matter. The second is heating by the enhancement
 of the reaction rate. It is apparent that in this case, after a small transient, the heating by enhanced reaction rates becomes dominant and the runaway takes place on a short timescale. For other underlying compositions the effect is a bit more complicated. 
 We discuss this issue in the next subsections.

\subsection{The Helium underlying layer}

\begin{figure} 
  \includegraphics[width=84mm]{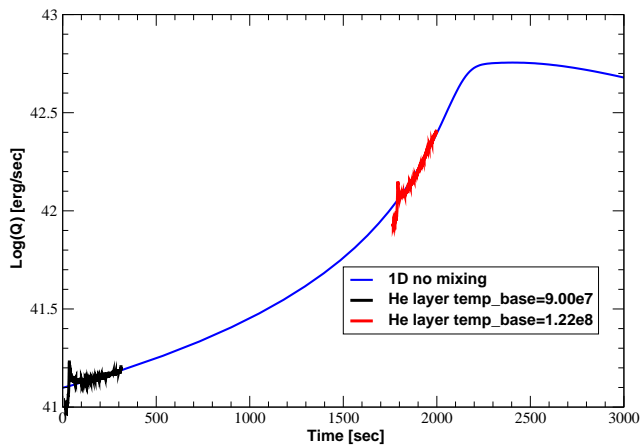}
  \caption{Log of the total energy production rate for all the helium models compared to the 1D model. The model with
   initial temperature higher than the default temperature of $ 9\times10^7 K $ was shifted in time by 1760.0 seconds.
   }
\hspace{1.0in}
  \label{Qhelium}
\end{figure}

\begin{figure}
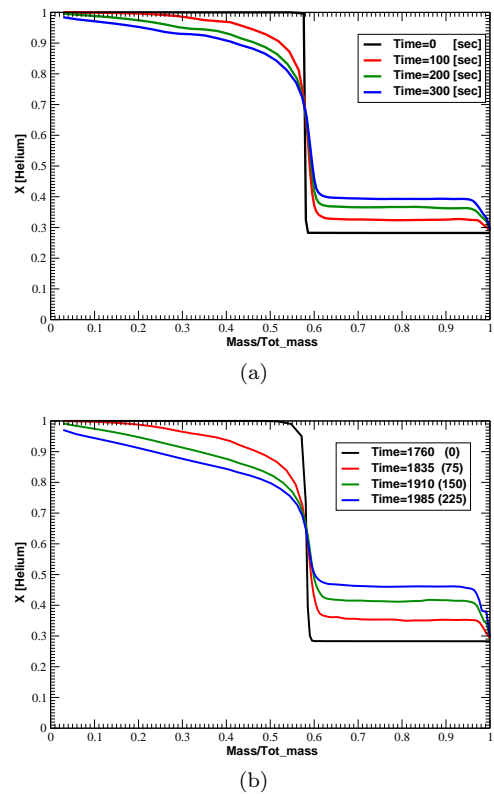

\centering

        \subfloat[]{\includegraphics[clip,width=.36\textwidth]{v-x-m12ag-dt100.eps}}

        \subfloat[]{\includegraphics[clip,width=.36\textwidth]{v-x-m12dg-dt75.eps}}

\caption{Abundance of the helium as function of the fractional mass coordinate 
of the computed envelope (see section 2) at various times that are indicated on the plot.
a- the model with the default base temperature of $9\times10^7 K$, 
b - the model with base temperature of $1.22\times10^8 K$ (times shifted by 1760 seconds, the 2D time is in brackets).
 } 
\label{Xhelium}
\end{figure}

 The helium enrichment, observed in recurrent nova (without enrichment by heavier isotopes) and in other
 classical nova was mentioned in the past as an obstacle to the underlying convective overshoot mechanism (\cite{lt87}).
 Helium is the most abundant end product of the hydrogen burning in nova outbursts and it does not enhance the 
 hydrogen burning in any way. In recurrent novae, helium may be accumulated upon the surface of the white dwarfs. 
 If so, can the dredge-up mechanism lead to the observed helium enrichment ?
 We examine this question using the case m12ag in  Table~\ref{tab:models}. Energetically, as expected,  the model follows
 exactly the 1D model (Fig.~\ref{Qhelium}). The slow rise of the burning rate in this case makes the 2D simulation too expensive.
 To overcome this problem we artificially  'jump in time' by jumping to another helium model at a later stage of the runaway, in which the 
 1D temperature is $1.22\times10^8 {\rm K}$. The rise time of this model is much shorter, and by adjusting 
 its time axis to that of the 1D model the two curves in Fig.~\ref{Qhelium} may be seen to coincide. The fluctuations of the 2D curve are absent in
 its 1D counterpart, both because the latter has no convection and because the 1D simulation is performed using implicit algorithm with
 much larger time steps. The precise way in which the 2D evolution agrees with the 1D evolution increases our confidence in the validity of the 2D simulations. 

\begin{figure*}
\centering
        \subfloat[][]{\includegraphics[width=.5\textwidth]{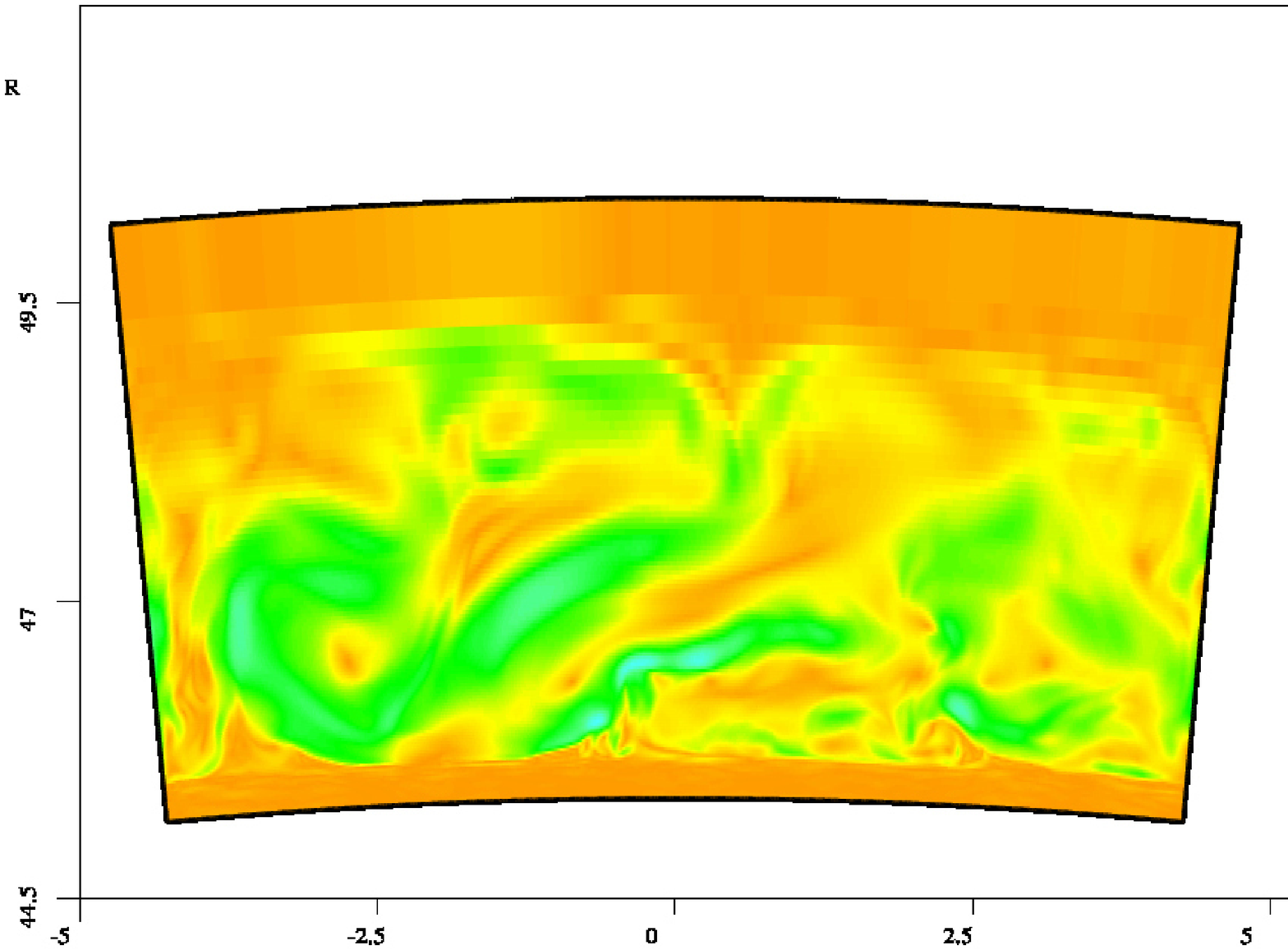}}
        ~
        \subfloat[][]{\includegraphics[width=.5\textwidth]{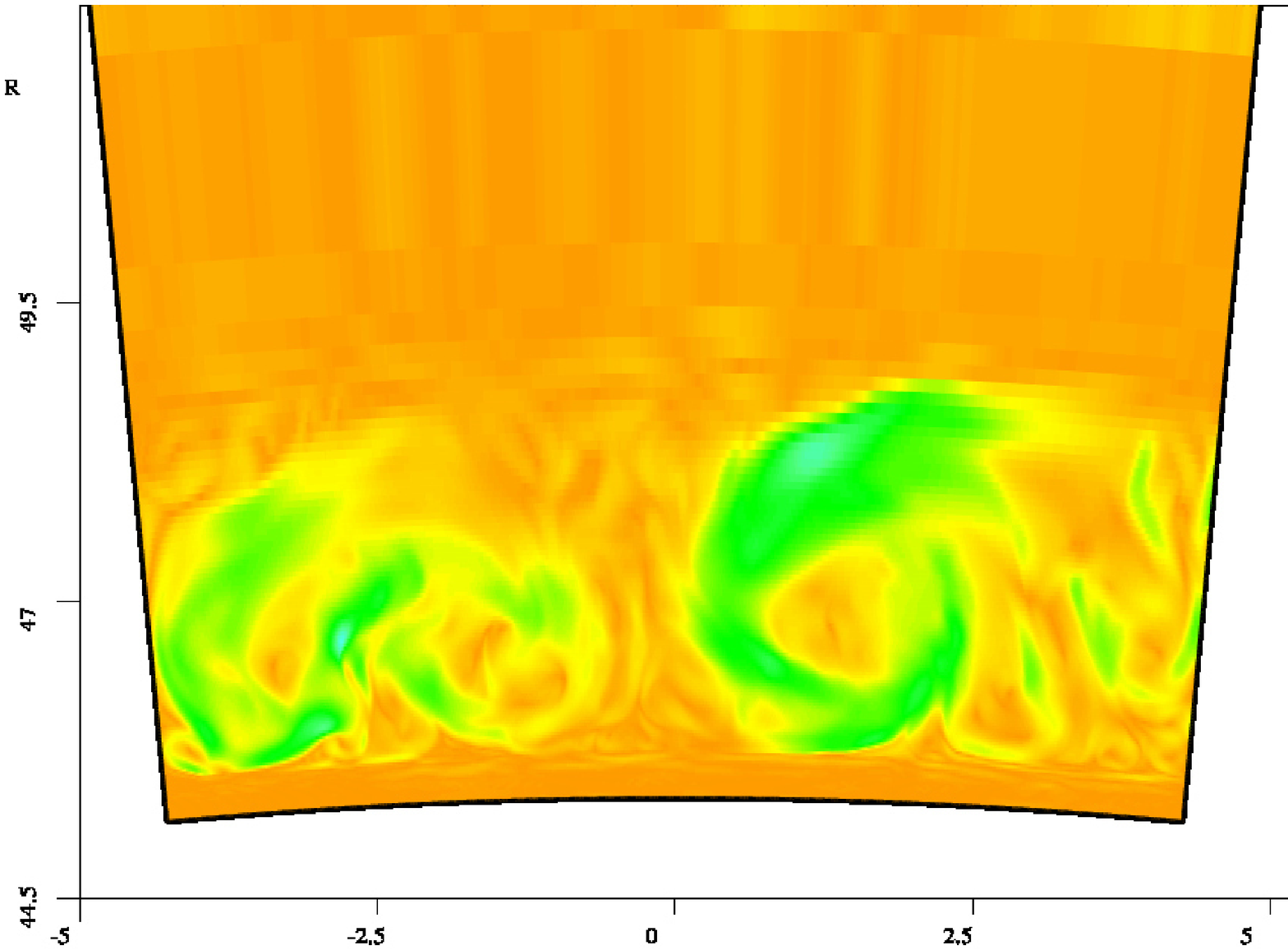}}

        \subfloat[][]{\includegraphics[width=.5\textwidth]{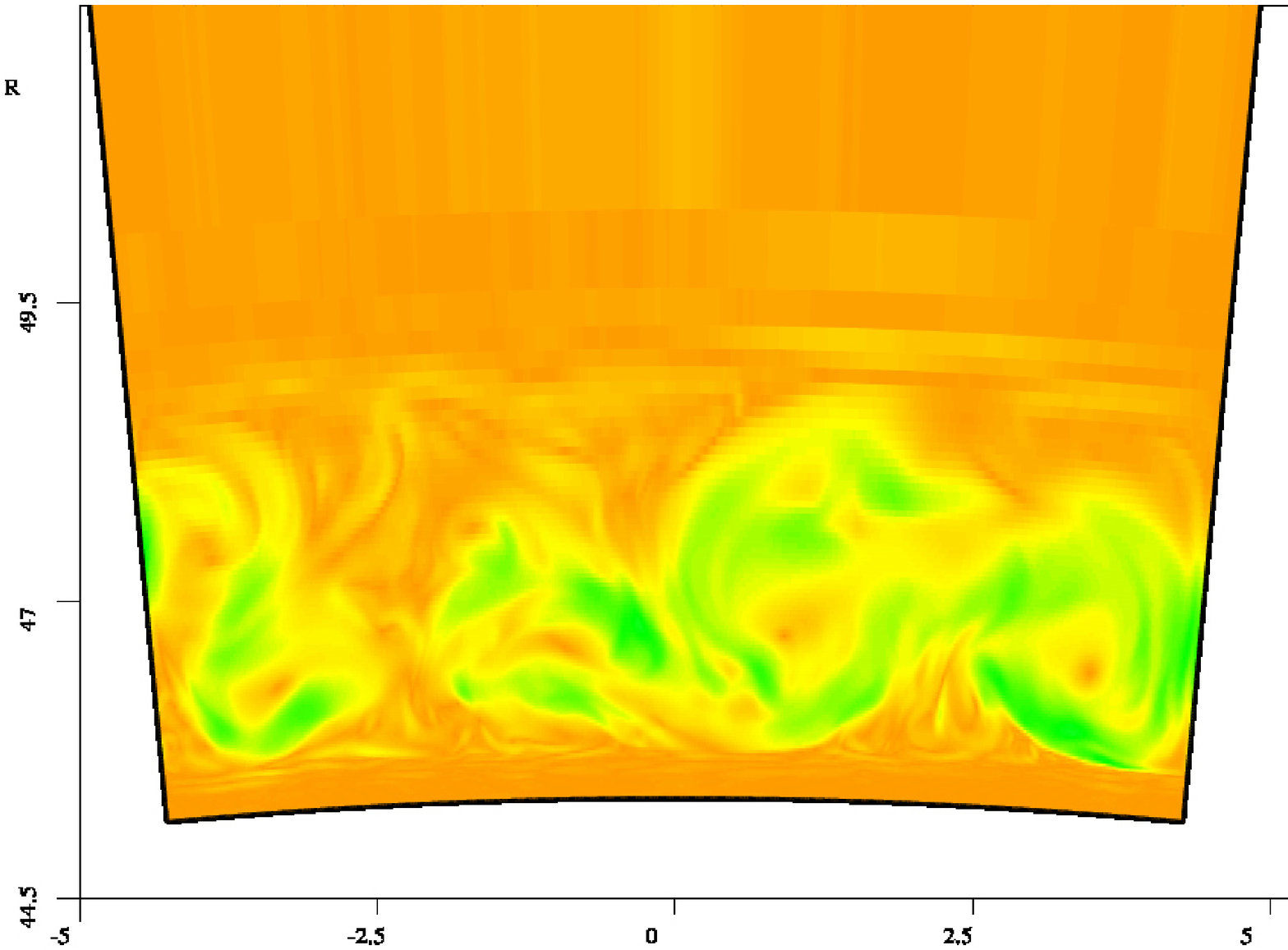}}
        ~
        \subfloat[][]{\includegraphics[width=.5\textwidth]{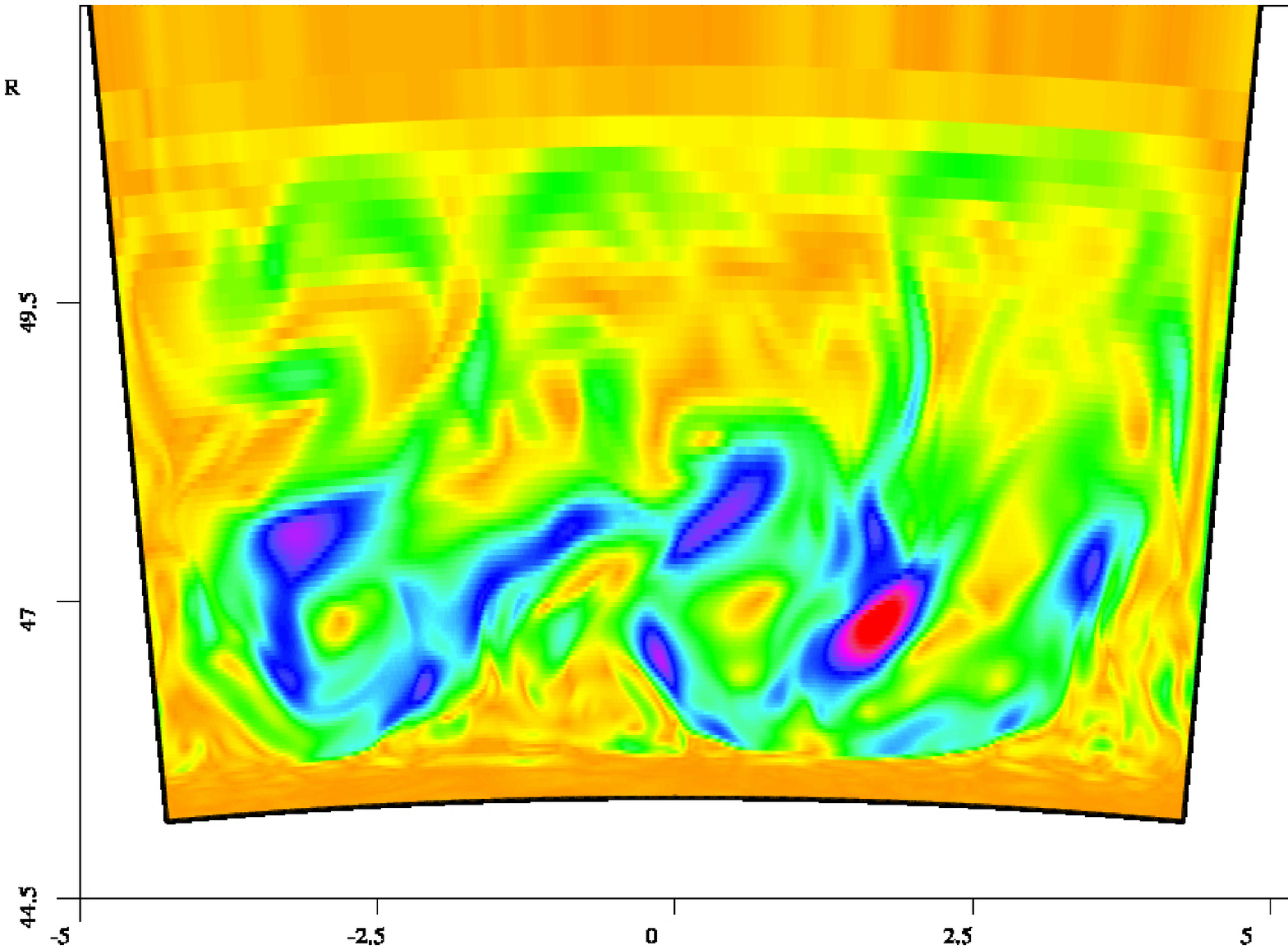}}

        \subfloat[][]{\includegraphics[width=.5\textwidth]{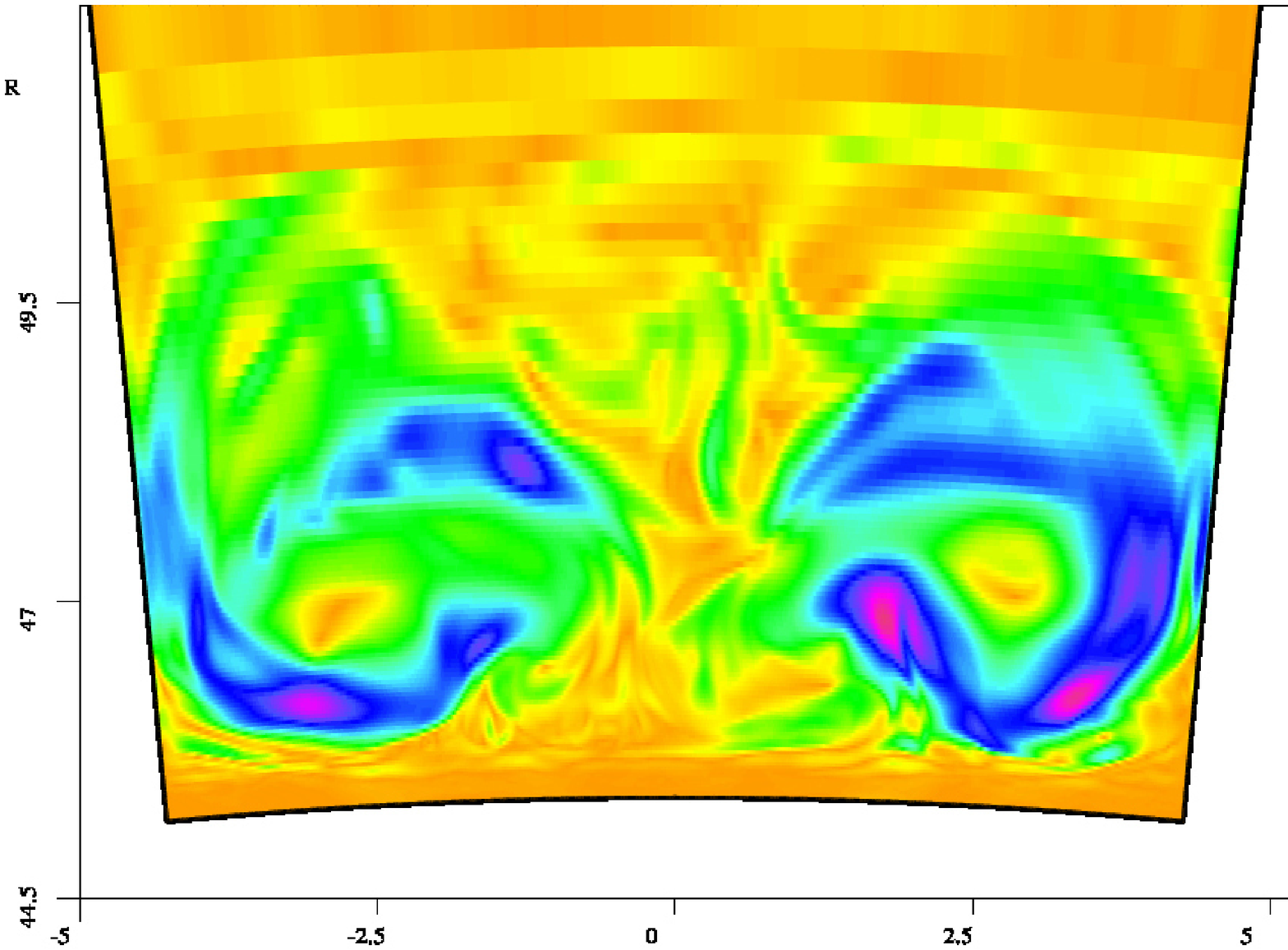}}
        ~
        \subfloat[][]{\includegraphics[width=.5\textwidth]{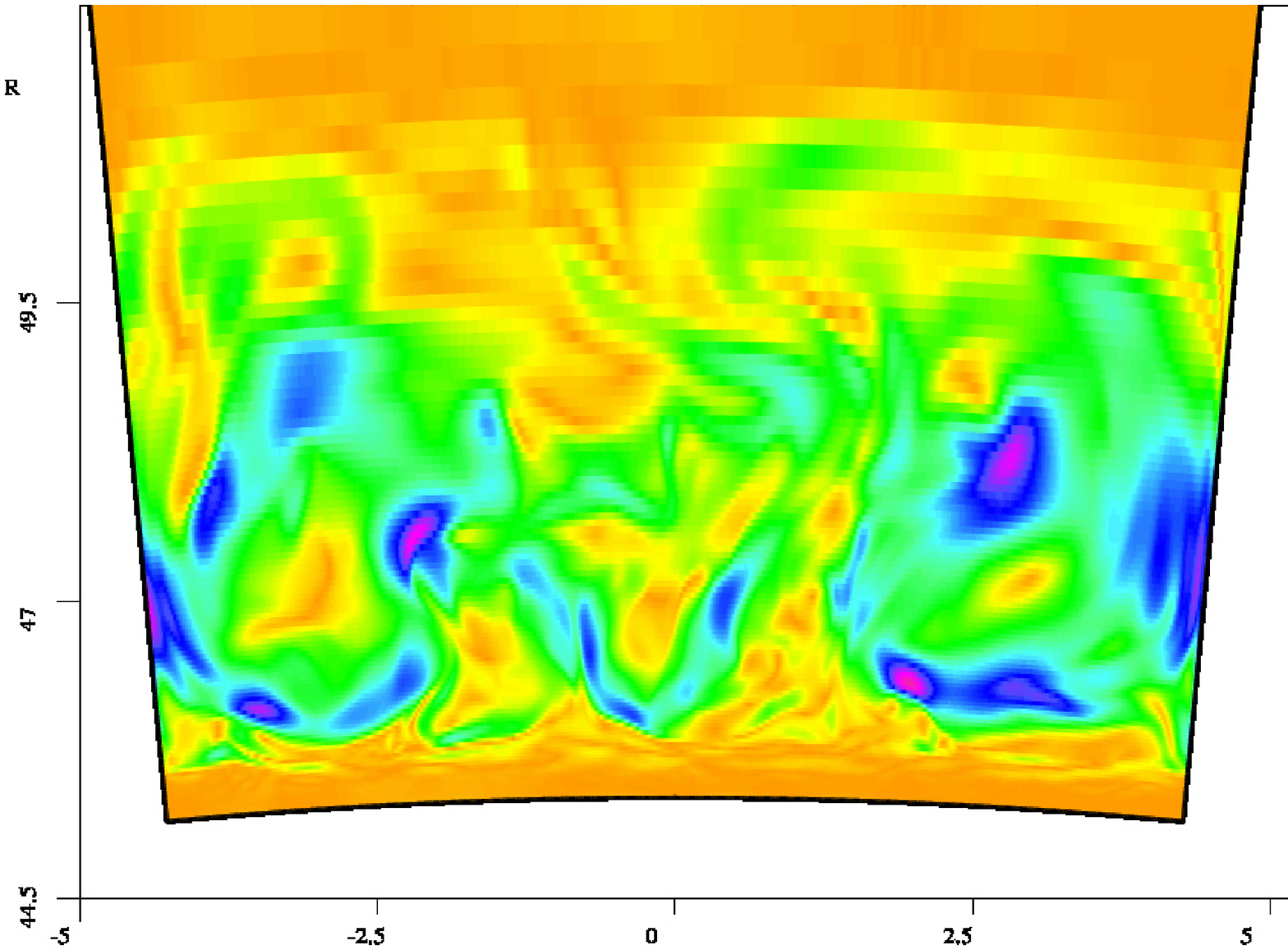}}
\caption{Color map of the convective speed for the underlying helium layer. The times are:
 model m12ag (a)- t=50 sec,(b)- t=200 sec, 
(c)- t=300 sec; model m12dg (d)- t=40 sec, (e)- t=140 sec , (f)- t=230 sec}
  \label{fig:flow-he}
\end{figure*}

   The convective flow is indeed moderate but overshoot mixing is observed at a certain level
 (Fig.~\ref{Xhelium}). In the first (slower) phase (m12ag in Table~\ref{tab:models}) the level of mixing is small,
 converging to about $10\%$. In the later (faster) phase (m12dg in Table~\ref{tab:models}) 
 the mixing rate increases in step with the increasing burning rate and the higher velocities in the convective cells. 
 However, as the rates are still low relative to the C-O case, the added amount of mixing is again about $10\%$, summing up to a total amount of about $20\%$.
 Since we begin with matter of solar composition, the total He mass fraction at the end of the second phase exceeds $40 \%$.   

 The color maps of the absolute value of the velocity in the 2D models at different times along the development of the runaway
 are presented in  Fig.~\ref{fig:flow-he}. In the first slower phase 
 (model m12ag Table~\ref{tab:models}), the burning rate is low and grows mildly with time. 
 The convective velocities are converging to a value of a few  $ 10^{6} $ cm/sec and the cell size is 
 only a bit larger than a scale height.
 In the second (faster) phase (model m12dg Table~\ref{tab:models}), the burning rate is somewhat higher and is seen to grow with time.
 The convective velocities are increasing with time up to a value of about  $1.5\times10^{7} $ cm/sec. The convective cells in the radial 
 direction converge with time to an extended  structure of a few scale heights.

\subsection{The ONe(Mg) underlying layer}

\begin{figure} 
  \includegraphics[clip,width=84mm]{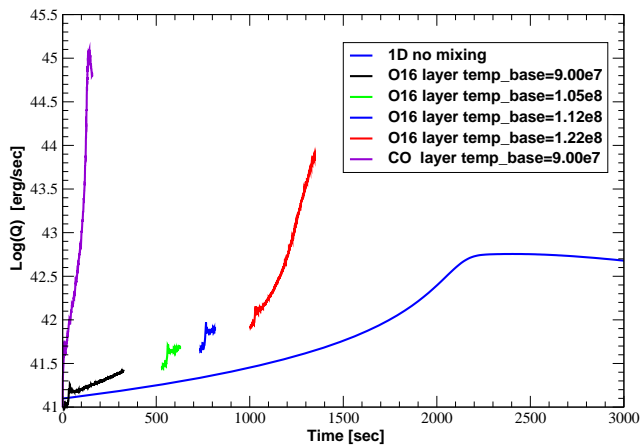}
  \caption{Log of the total energy production rate for all the oxygen models compared to the 1D model and to the CO model. 
   The models with initial temperature higher than the default temperature of $ 9\times10^7 K $ were shifted in time in order 
   to produce a smooth continuous line.}
  \label{Qoxygen}
\end{figure}

 The rate of proton capture by oxygen is much slower and less energetic than the capture
 by carbon but it still has an enhancing effect relative to the 1D model without mixing (Fig.~\ref{Qoxygen}). 
 This can be well overstood once we notice that for the initial temperature of the 2D model i.e.  $ 9.0\times 10^7 K$  
 the energy generation rate by proton capture by oxygen is more than three orders of magnitude lower than the energy
 generation rate by carbon capture (Fig.~\ref{fig:Qcapture}). We can also observe that the energy generation rate for capture 
 by oxygen stays much smaller than the energy generation rate for capture by carbon for the entire range of temperature
 relevant to nova outbursts. For this range, we also notice that the capture by \iso{20}Ne is lower by an order of magnitude than
 the capture rate by \iso{16}O. 
 Being interested only in the energetics, convective flow and mixing by dredge up, we choose to seperate varibles and study the case of 
 a pure  \iso{16}O underlying layer as a test case (ignoring any possible \iso{24}Mg that is predicted by evolutionary codes). 

 We computed the model for an integrated time of 300 seconds (about one million time steps). The trend is very clear
 and we can extrapolate and predict a runaway a bit earlier and more energetic than the 1D case. Again, we could not continue
 that 2D simulation further due to the very low burning rates and a small hydrodynamical time step. As before, we computed three
 different phases, where each of them starts from a different 1D model along its evolution. The maximal 1D temperatures at the base of the burning shell,
 when mapped to the 2D grid, are : $ 1.05\times 10^8 K$, $ 1.125\times 10^8 K$  and $ 1.22\times 10^8 K$  respectively (Table~\ref{tab:models}).
 All three phases start with a transient related to the buildup of the convective flow. In Fig.~\ref{Qoxygen} we shifted the curves of burning rates
 in time in a way that permits a continuous line
 to be drawn. Along this continuous line, evolution proceeds faster towards a runaway than the 1D burning rate, also shown in Fig.~\ref{Qoxygen}.

\begin{figure} 
  \includegraphics[clip,width=84mm]{Qcapture.eps}
  \caption{Log of the energy generation rate for proton capture on: \iso{12}C,  \iso{16}O,  \iso{20}Ne, and \iso{24}Mg, the rate is calculated for $\rho=1000.0$ gr/cc.}
  \label{fig:Qcapture}
\end{figure}

\begin{figure}
\centering
        \subfloat[][]{\includegraphics[clip,width=.36\textwidth]{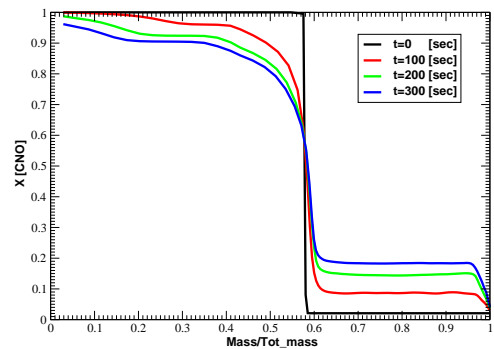}}
        
        \subfloat[][]{\includegraphics[clip,width=.36\textwidth]{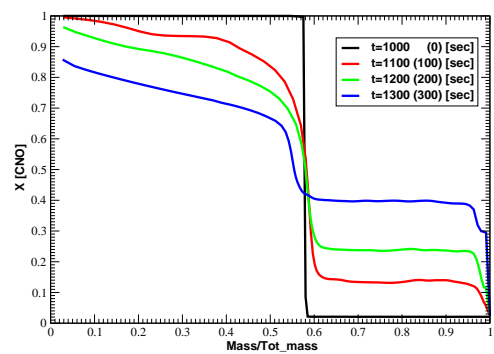}}

\caption{Abundance of the CNO elements as function of the fractional mass 
coordinate of the computed envelope (see section 2) at various times that are indicated on the plot.
a - the model with the default base temperature of $9\times10^7 K$, 
b - the model with base temperature of $1.22\times10^8 K$.
 } 
\label{Xoxygen}
\end{figure}

 We computed the last phase, with an initial base temperature of $ 1.22\times 10^8 K$ , for 350 sec until it reached a maximum.
 The maximal achieved temperature is $ 2.15\times 10^8 K$  and the maximal achieved burning rate is  $ 82.4\times 10^{42} erg/sec $. As expected, this case lies somewhere between the 1D case and the C-O model.
 The convective flow resembles the C-O case, a significant feature being the strong correlation between the burning rate and the
 convective velocities. Most importantly for the case of an underlying oxygen layer, dredge up of substantial amounts of matter from the core into the envelope
 occurs in all our simulations. The trends are about the same as for the C-O models (Fig.~\ref{Xoxygen}).
 The correlation of the amount of mixing with the intensity of the burning is easily observable.

\subsection{Approximate models for \iso{24}Mg underlying layer}
\label{mgreac}

\begin{figure} 
  \includegraphics[width=84mm]{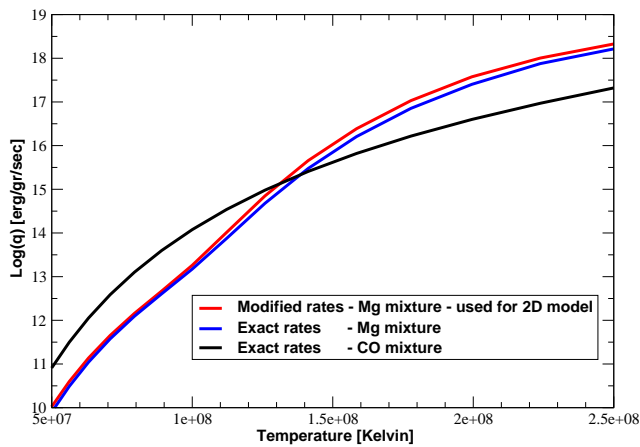}
  \caption{Log of the total energy production rate for $\rho=1000.0$ gr/cc of a mixture that contains $90\%$ solar matter
    mixed with $10\%$ of \iso{24}Mg.
   Red:15 elements net used for the 2D model; Blue: The rates given by a full net of 216 elements. The Black line gives the
   rates of  $90\%$ solar matter mixed with $10\%$ of CO core matter (see text).
   }
  \label{fig:Mg-rates}
\end{figure}

Nova outbursts on massive ONe(Mg) white dwarfs are expected to be energetic fast nova \citep{sst86,Gehrz98,Gehrz08,Ili02}.
The problem we face is whether,  in absence of the enhancement by  \iso{12}C, the overshoot mixing mechanism
can generate such energetic outbursts by mixing the solar abundance accreted matter with the underlying ONe(Mg) core.
Based upon examination of the energy generation rate of proton capture reactions $(p,\gamma)$ on
 \iso{12}C,  \iso{16}O,  \iso{20}Ne
, and \iso{24}Mg the results shown in Fig.~\ref{fig:Qcapture} make it evident that only \iso{24}Mg  can compete with \iso{12}C
 in the range of temperatures relevant to nova runaways. Therefore, in spite of the fact that the abundance of \iso{24}Mg in the core sums up to only a few percent (\cite{GpGb01,siess06}),
the high capture rate might compensate for the low abundances and play an important role in the runaway.
Furthermore, previous studies show that in the outer parts of the core, the parts important for our study, \iso{24}Mg is more abundant and can represent up to about $25 \%$ (\cite{berro94}).                  
 Restricting ourselves to the reaction network that includes only 15 elements 
 we assume, as a demonstration, an artificial case of a homogeneous underlying layer with only one isotope (\iso{24}Mg).
 For this homogeneous layer model we replaced the energy generation rate
 of proton capture by  \iso{16}O with the values of proton capture by \iso{24}Mg Fig.~\ref{fig:Qcapture}. 
 To check our simplified network, we present in Fig.~\ref{fig:Mg-rates} the rates computed by a 216 elements network and our modified 15 elements rates,
 both for  a mixture of $90\%$ solar matter with $10\%$ \iso{24}Mg. The difference is much smaller than the difference inside the big network between this mixture and a mixture of $90\%$ solar matter with $10\%$ C-O core matter. Therefore our simplified network is a good approximation regarding energy production rates. 

 The crossing of the curves in Fig.~\ref{fig:Mg-rates} reveals a striking and very important feature - at temperatures less than $1.3\times10^{8}$K a mixture of $10\%$ carbon and
 $90\%$ solar compostion burns roughly ten times faster than a mixture of $10\%$ magnesium with the same solar composition gas. Above that temperature the
 rates exchange places and magnesium enhancement dominates C-O enhancement. This emphasizes the importance of a proper treatment of the effects of the $^{24}$Mg abundance in explosive burning on ONe(Mg) white dwarfs.
 In a future work, we intend to study more realistic models with the inclusion of a detailed reaction network.

\begin{figure} 
  \includegraphics[width=84mm]{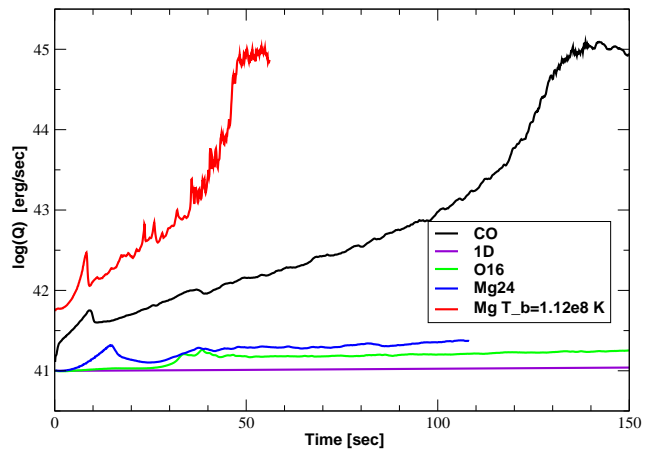}
  \caption{Log of the total energy production rate for all the magnesium models compared to the CO model the 1D model and the oxygen model. 
  The model with initial temperature higher than the default temperature of $ 9\times10^7 K $ is not shifted in time (see text).
   }
  \label{Qmagnesium}
\end{figure}

 In Fig.~\ref{Qmagnesium} we present the total burning rates in our toy model, together with the rates of previous models.

 As can be expected, the enhancement of the burning in this toy model, relative to the 1D model, is indeed observed. 
 However, the development of the runaway although faster than in
 the underlying  \iso{16}O  model is still much slower  than the rise time of the C-O model.
 This result is easily overstood via the discussion above, as the initial burning temperature in the model is only ( $9\times10^7 {\rm K}$). At that temperature, the energy 
 generation rate for proton capture on \iso{24}Mg is lower by almost three orders of magnitude relative to the energy generation rate by proton capture on
 \iso{12}C. The rates are about the same when the temperature is  $1.3\times10^8 {\rm K}$ and from there on the
 magnesium capture rate is much higher. In order to demonstrate that this is indeed the case we calculated another 2D magnesium model
 in which the initial maximal 1D temperature was $1.125\times10^8 {\rm K}$. The rise time of this model is very short, even shorter than 
 the rise time of the C-O model. One should regard those two simulations as two phases of one process - slow and fast.
 The maximal achieved temperature is $ 2.45\times 10^8$K   
 and the maximal achieved burning rate is  $ 1000.0\times 10^{42} erg/sec $, similar to the C-O case.

\begin{figure*}
\centering
        \subfloat[][]{\includegraphics[width=.5\textwidth]{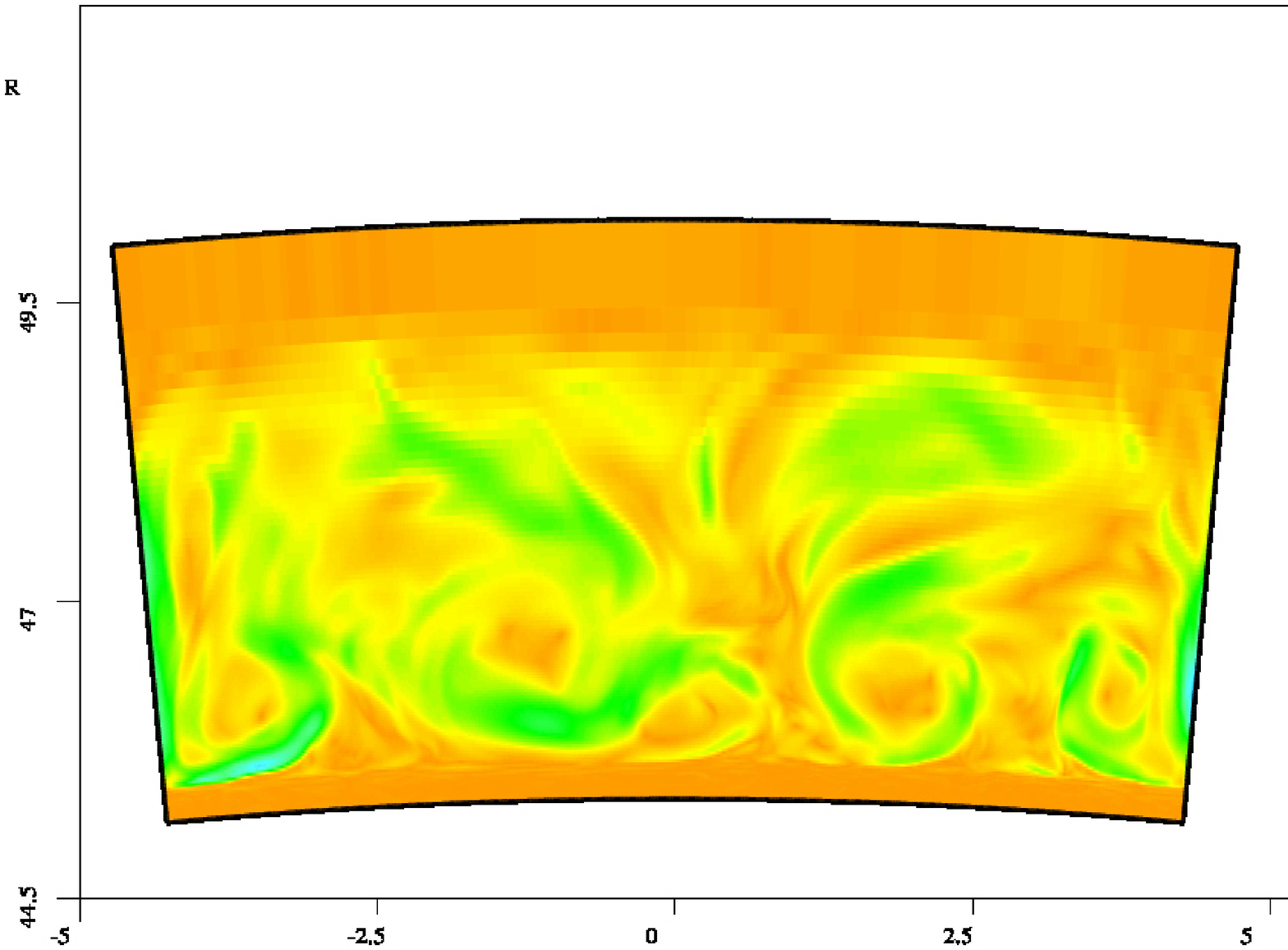}}
        ~
        \subfloat[][]{\includegraphics[width=.5\textwidth]{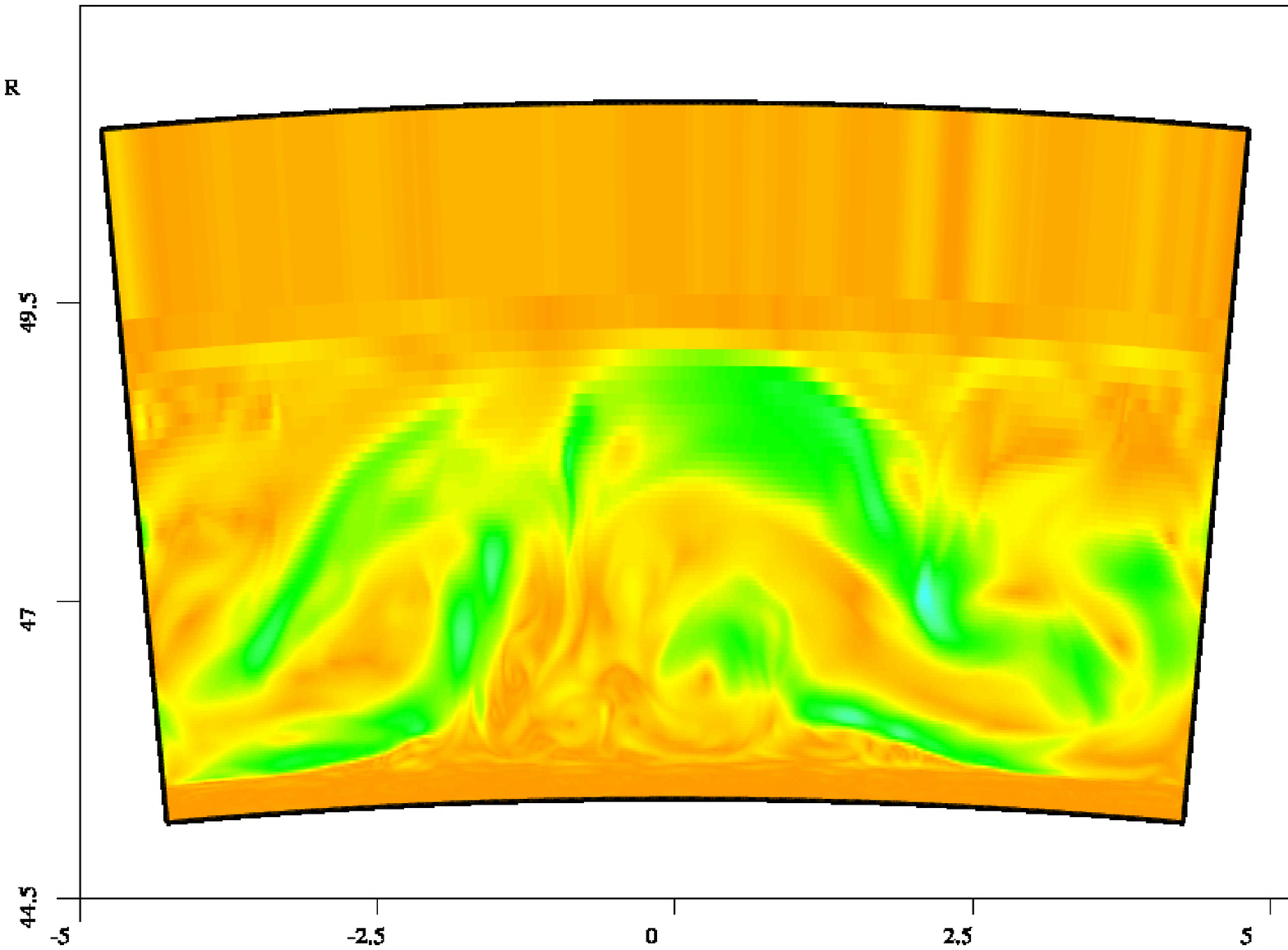}}

        \subfloat[][]{\includegraphics[width=.5\textwidth]{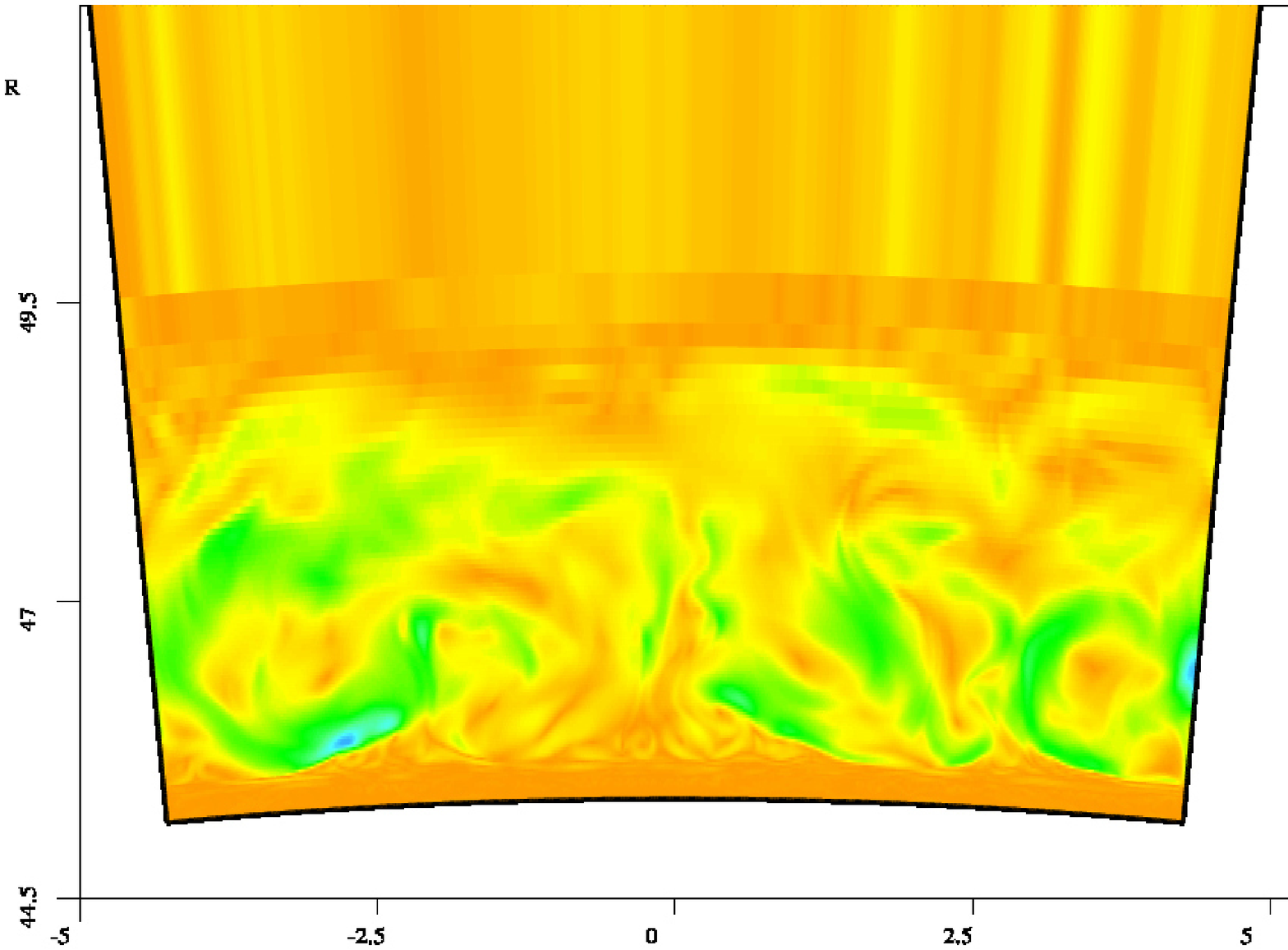}}
        ~
        \subfloat[][]{\includegraphics[width=.5\textwidth]{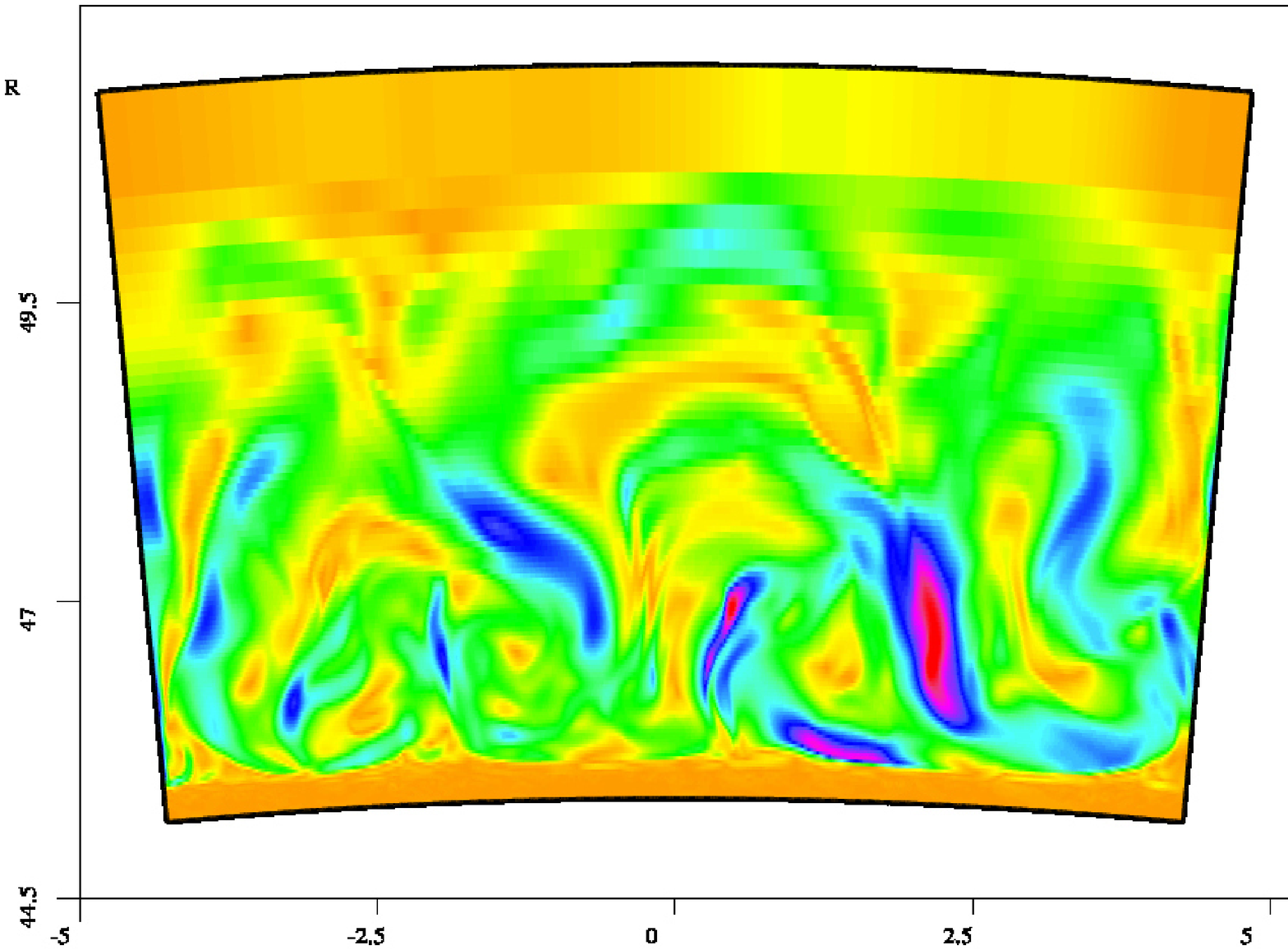}}

        \subfloat[][]{\includegraphics[width=.5\textwidth]{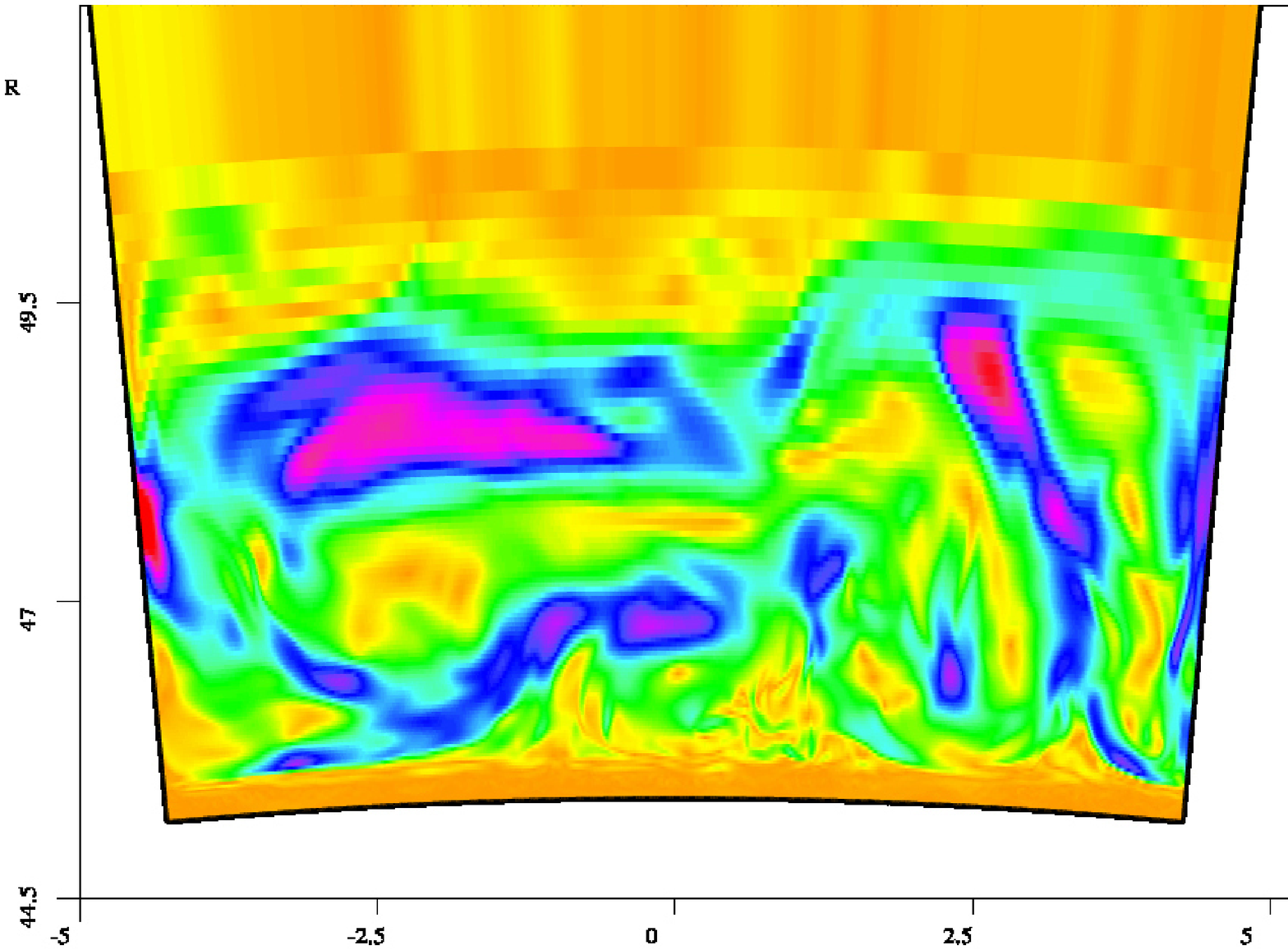}}
        ~
        \subfloat[][]{\includegraphics[width=.5\textwidth]{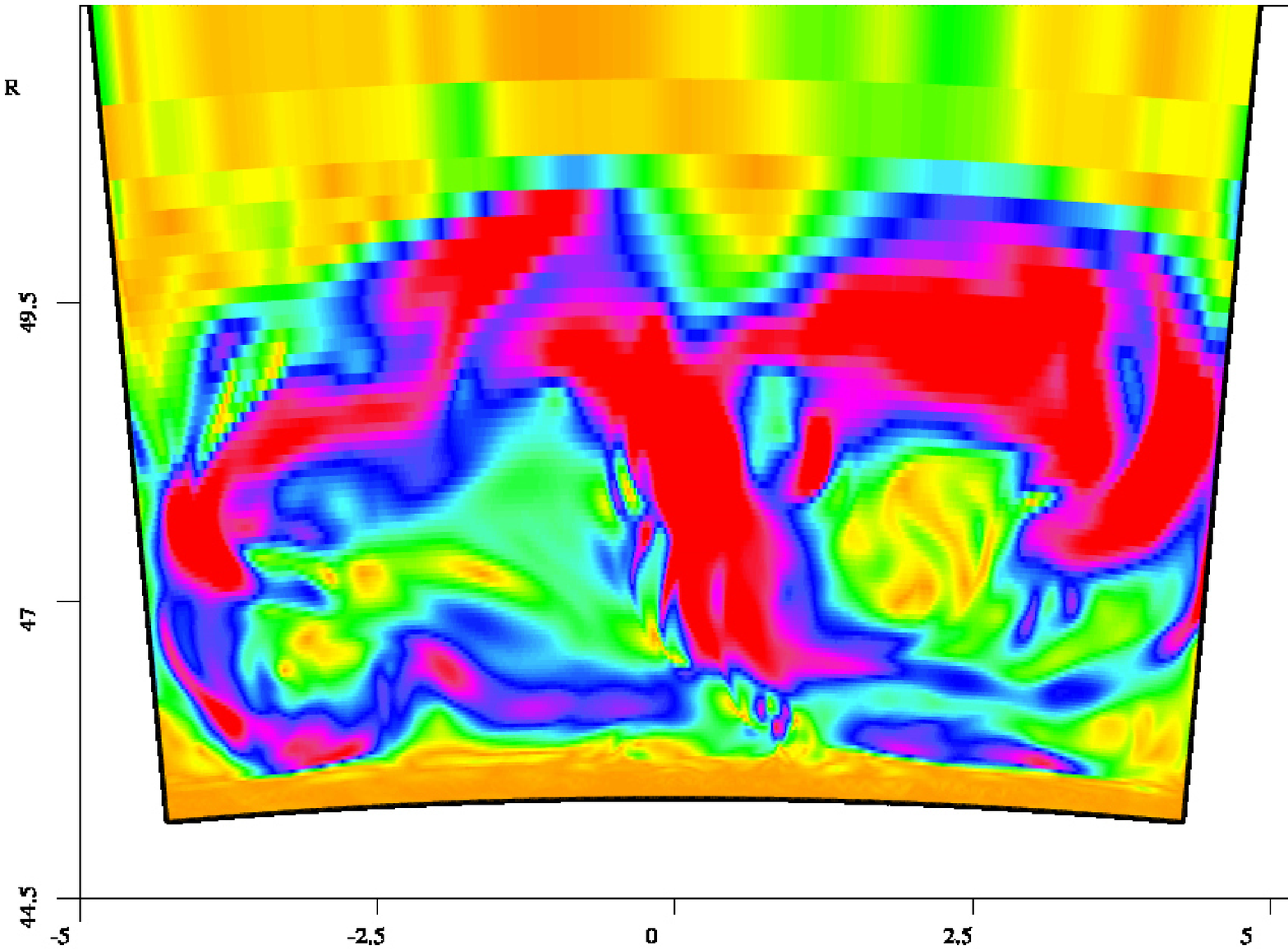}}
\caption{Color map of the convective speed for the underlying magnesium layer. The times are:
 model m12kk (a)- t=40 sec, (b)- t=60 sec, (c)- t=70 sec; model m12jj (d)- t=20 sec, (e)- t=30 sec ,(f)- t=40 sec
 }
  \label{fig:flow-mg}
\end{figure*}

  In order to better understand the convective flow for the case with  underlying \iso{24}Mg we generated color maps of the
 absolute value of the velocity (speed) in the 2D models at different times along the development of the runaway
  (Fig.~\ref{fig:flow-mg}). The two \iso{24}Mg cases show extremely different behavior. In the first case 
 (model m12kk Table~\ref{tab:models}), the burning rate is low and it grows mildly with time. 
 The convective velocities are converging to a value of a few  $ 10^{6} $ cm/sec and the cell size is 
 only a bit bigger than a scale height.
 In the second case  (model m12jj Table~\ref{tab:models}), the burning rate is high and it grows rapidly with time.
 The convective velocities are increasing with time  up to a value of a few  $10^{7} $ cm/sec. The convective cells in the radial 
 direction converge to a structure of a few scale heights.


\begin{figure}
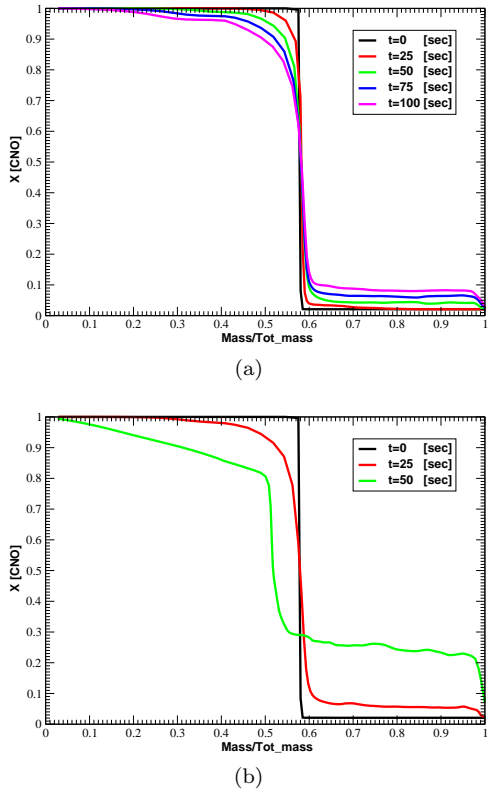

\centering
        \subfloat[][]{\includegraphics[clip,width=.36\textwidth]{v-x-m12kk-dt25.eps}}
        
        \subfloat[][]{\includegraphics[clip,width=.36\textwidth]{v-x-m12jj-dt25.eps}}
\caption{Mixing for the underlying magnesium layer,abundance of the CNO elements  as function of the fractional mass 
coordinate of the computed envelope (see section 2) at various times that are indicated on the plot.
a - the model with the default base temperature of $9\times10^7 K$, 
b - the model with base temperature of $1.125\times10^8 K$.
 } 
\label{Xmagnesium}
\end{figure}
In accordance with our previous cases, the magnesium toy model dredge up substantial amounts of matter from the core to the envelope.
 There is a one to one correlation with the convective velocities. The amount of mixing at the slow initial stages
 (model m12kk) are small and tend to converge to a few percents. The amount of mixing at the late fast stages
 grows rapidly with time (Fig.~\ref{Xmagnesium}). We present here only the general trend detailed results will be presented 
 in a forthcoming study.

 \section{conclusions}
 \label{conclu}

 We present here, for the first time, detailed 2D modeling of nova eruptions for a range of possible compositions beneath the accreted hydrogen layer.
 The main conclusion to be drawn from this study is that \bf {significant enrichment (around $30 \%$) of the ejected layer, by the convective drege-up mechanism, is a common feature of the
 entire set of models, regardless of the composition of the accreting white dwarf} \rm. On the other hand, the burning rates and therefore the time scales of the runaway
 depend strongly on the composition of the underlying layers. There is also a one to one correlation between the burning rate, the velocities in the convective
 flow, and the amount of temporal mixing. Therefore, second order differences in the final enrichment are expected to depend on the underlying composition.
 Specific results for each case are as follows :
      
     a)  Since the energy generation rate for the capture of protons by \iso{12}C is high for  the entire temperature range
         prevailing both in the ignition of the runaway and during the runaway itself, the underlying carbon layer
         accelerates the ignition and gives rise to C-O enrichment in the range of the observed amounts. 

     b)  For the densities and temperatures prevailing in nova outbursts helium is an inert isotope. 
         Therefore, it does not play any role in the enhancement of the runaway. Nevertheless, we demonstrate that 
         once the bottom of the envelope is convective, the shear flow induces substantial amounts of mixing with the 
         underlying helium. The eruption in those cases is milder, with a lower burning rate. For recurrent nova, 
         where the timescales are too short for the diffusion process to play a significant role, the observed helium 
         enrichment  favor the underlying convection mechanism as the dominant mixing mechanism. Future work dealing 
         with more realistic core masses (1.35-1.4 solar masses) for recurrent novae will give better quantitative predictions 
         that will enable us to confront our results with observational data. 

     c)  The energy generation rate for the capture of a proton by \iso{16}O is much lower than that of the capture by \iso{12}C for the 
         entire temperature range prevailing in the ignition of the runaway and during the runaway itself. Underlying oxygen, 
         whenever it is present, is thus expected to make only a minor contribution to the enhancement of the runaway.
         As a result the time scale of the runaway in this case is much larger than that of the C-O case. Still, the final enrichment
         of the ejecta is above $40 \% $, (Fig.~\ref{Xoxygen}). The energy generation rate by the capture of a proton by \iso{20}Ne is even lower than that
         of capture rate by \iso{16}O. We therefore expect \iso{20}Ne to make again only a minor contribution to the enhancement of the runaway, but with
         substantial mixing.

     d)  Nova outbursts on massive ONe(Mg) white dwarfs are expected to be energetic fast nova. In this survey we show that for 
         the range of temperatures relevant for the nova runaway the only isotope that can compete with the  \iso{12}C as a source 
         for burning enhancement by overshoot  mixing is \iso{24}Mg. From our demonstrating toy model, we can speculate that even
         small amounts of \iso{24}Mg present at the high stages of the runaway can substantially enhance the burning rate, 
         leading to a faster runaway with a significant amount of mixing. The relationship between the amount of \iso{24}Mg in the ONe(Mg) core,
         the steepness of the runaway, and the amount of mixing in this case are left to future studies.

\section{Acknowledgments} 

We thank the referee for his comments which helped us in clarifying
our arguments in the revised version of the paper.
Ami Glasner, wants to thank the Department of Astronomy and Astrophysics at 
the University of Chicago for the kind hospitality during his visit to Chicago, 
where part of this work was done.
This work is supported in part at the University of Chicago by the National Science Foundation under
Grant PHY 02-16783 for the Frontier Center "Joint Institute for Nuclear Astrophysics" (JINA), and in 
part at the Argonne National Laboratory by the U.S. Department of Energy, Office of Nuclear Physics,
under contract DE-AC02-06CH11357.

\label{lastpage}

\end{document}